%% file: main.tex
\documentclass{SIP}

\usepackage{amsmath}
\usepackage{hyperref}
\usepackage{graphics}
\usepackage{algorithmic}
\usepackage{subfig}
\usepackage{url}
\usepackage{epstopdf, epsfig} 
\usepackage{balance}

\begin{document}

\title[The Age of Synthetic Realities]{\centering{The Age of Synthetic Realities:}\\\centering{Challenges and Opportunities}\\}
\author[Cardenuto, \textit{et al}.]{Jo\~{a}o~Phillipe~Cardenuto$^{1}$, Jing~Yang$^{1}$, Rafael~Padilha$^{1}$, Renjie~Wan$^{2}$, Daniel~Moreira$^{3}$, Haoliang~Li$^{4}$, Shiqi~Wang$^{5}$, Fernanda~Andal\'{o}$^{1}$, S\'{e}bastien~Marcel$^{6,7}$ and Anderson~Rocha$^{1}$}


\address{\add{1}{Artificial Intelligence Lab., \url{Recod.ai}, Institute of Computing, University of Campinas, Campinas, SP, Brazil.}
\add{2}{Department of Computer Science, Hong Kong Baptist University, Hong Kong.}
\add{3}{Department of Computer Science, Loyola University Chicago, Chicago, IL, USA.}
\add{4}{Department of Electrical Engineering, City University of Hong Kong, Hong Kong.}
\add{5}{Department of Computer Science, City University of Hong Kong, Hong Kong.}
\add{6}{Idiap Research Institute, Martigny, Switzerland.}
\add{7}{University of Lausanne, Lausanne, Switzerland.}
}

\corres{\name{Jo\~{a}o~Phillipe~Cardenuto}
\email{phillipe.cardenuto@ic.unicamp.br}}

\begin{abstract}
Synthetic realities are digital creations or augmentations that are contextually generated through the use of Artificial Intelligence (AI) methods, leveraging extensive amounts of data to construct new narratives or realities, regardless of the intent to deceive. In this paper, we delve into the concept of synthetic realities and their implications for Digital Forensics and society at large within the rapidly advancing field of AI. We highlight the crucial need for the development of forensic techniques capable of identifying harmful synthetic creations and distinguishing them from reality. This is especially important in scenarios involving the creation and dissemination of fake news, disinformation, and misinformation. Our focus extends to various forms of media, such as images, videos, audio, and text, as we examine how synthetic realities are crafted and explore approaches to detecting these malicious creations. Additionally, we shed light on the key research challenges that lie ahead in this area. This study is of paramount importance due to the rapid progress of AI generative techniques and their impact on the fundamental principles of Forensic Science.
\end{abstract}


\maketitle

\section{Introduction}
\input{sections/intro}

\section{Synthetic Images}
\input{sections/syn-images}

\section{Synthetic Videos}
\input{sections/syn-videos}
\section{Synthetic Audio}
\input{sections/syn-audio}

\section{Synthetic Text}
\input{sections/syn-text}
\section{NERFs and Metaverse}
\input{sections/metaverse}
\section{DeepFakes}
\input{sections/deepfakes}


\section{Conclusion and Final Thoughts}
\input{sections/challenges}

\section*{Acknowledgement}
The Swiss, Brazilian and U.S. Governments are authorized to reproduce and distribute reprints for Governmental purposes, notwithstanding any copyright notation thereon. The views and conclusions contained herein are those of the authors and should not be interpreted as necessarily representing the official policies or endorsements, either expressed or implied, of Swiss and Brazil governments or of DARPA, AFRL or the U.S. Government.

\section*{Financial Support}
This work was supported by the S\~{a}o Paulo Research Foundation (FAPESP) (J.P.C., grant numbers 2020/02211-2, 2017/12646-3), (J.Y., grant numbers 2019/04053-8, 2022/05002-0), (A.R., grant number 2017/12646-3); Blue Sky Research Fund of HKBU (R.W., grant number BSRF/21-22/16); Guangdong Basic and Applied Basic Research Foundation (R.W., grant number 2022A15151106\\92); Defense Advanced Research Projects Agency (DARPA) and the Air Force Research Laboratory (AFRL) (D.M., grant number FA8750-20-2-1004); and the Research Grant Council (RGC) of Hong Kong through Early Career Scheme (ECS) (H.L., grant number 21200522). Also, part of this work within the Idiap Research Institute received funding from the Swiss State Secretariat for Education, Research and Innovation (SERI), the Swiss State of Valais and the City of Martigny.

\bibliographystyle{IEEEtran}
\balance
\bibliography{main.bib}

\noindent \large \textbf{Biographies} 
\vskip2pc

\noindent\normalsize\textbf{Jo\~{a}o~Phillipe~Cardenuto} received a degree in Bachelor in Computer Engineering and Computer Science at the University of Campinas (2019). Currently, he is pursuing a Ph.D. degree at the University of Campinas, Brazil. Due to his research work, he was honored with the Google Latin America Research Award (LARA) 2021. He also had the opportunity to collaborate with Google Research as an intern in 2022. His latest works have focused on image provenance analysis and the development of forensic algorithms aimed at detecting doctored scientific images. His research interests include media forensics, computer vision, machine learning, and scientific integrity.

\vskip2pc

\noindent\textbf{Jing Yang} is a Ph.D. student of the Artificial Intelligence Lab., Recod.ai, and is currently doing a research internship at the Ubiquitous Knowledge Processing (UKP) Lab. She received her Master’s degree in Computer Science at Hunan University, China, and her Bachelor’s in Information and Computing Science at Hubei University of Technology, China. Jing’s research interests include natural language understanding, fact-checking, and forensics.

\vskip2pc

\noindent\textbf{Rafael Padilha} is a researcher associated with the Artificial Intelligence Lab. (Recod.ai) at the Institute of Computing, University of Campinas, Brazil. He received his Ph.D. in 2022 from the same university, with a joint research internship at the University of Kentucky, USA. He works for Microsoft Research, innovating on agriculture and sustainability under the Research for Industry team. His research interests lie in computer vision, machine learning, and digital forensics.

\vskip2pc

\noindent\textbf{Renjie Wan} received his BEng degree from the University of Electronic Science and Technology of China in 2012 and the Ph.D. degree from Nanyang Technological University, Singapore, in 2019. He is currently an Assistant Professor at Hong Kong Baptist University, Hong Kong. He is the outstanding reviewer of ICCV 2019 and the recipient of the Microsoft CRSF Award, VCIP 2020 Best Paper Award, and the Wallenberg-NTU Presidential Postdoctoral Fellowship.

\vskip2pc

\noindent\textbf{Daniel Moreira} received a Ph.D. degree in computer science from the University of Campinas, Brazil, in 2016. After working four years as a systems analyst with the Brazilian Federal Data Processing Service (SERPRO), he joined the University of Notre Dame for six years, first as a post-doctoral fellow and later as an assistant research professor. He is currently an assistant professor in the Department of Computer Science at Loyola University Chicago. He is also a member of the IEEE Information Forensics and Security Technical Committee (IFS-TC), 2021-2023 term, IEEE Signal Processing Society Education Center Editorial Board, 2022-2023 term, and associate editor of IEEE Transactions on Information Forensics and Security (T-IFS) and Elsevier Pattern Recognition journals. His research interests include media forensics, machine learning, computer vision, and biometrics.

\vskip2pc

\noindent\textbf{Haoliang Li} received his Ph.D. degree from Nanyang Technological University (NTU), Singapore in 2018. He is currently an assistant professor in Department of Electrical Engineering, City University of Hong Kong. His research mainly focuses on AI security, multimedia forensics and transfer learning. He received the Wallenberg-NTU presidential postdoc fellowship in 2019, doctoral innovation award in 2019, VCIP best paper award in 2020, Top 50 Chinese Young Scholars in AI+X 2022, and Stanford's top 2\% most highly cited scientists in 2022.

\vskip2pc

\noindent\textbf{Shiqi Wang} received the B.S. degree in computer science from the Harbin Institute of Technology in 2008 and the Ph.D. degree in computer application technology from Peking University in 2014. From 2014 to 2016, he was a Post-Doctoral Fellow with the Department of Electrical and Computer Engineering, University of Waterloo, Waterloo, ON, Canada. From 2016 to 2017, he was a Research Fellow with the Rapid-Rich Object Search Laboratory, Nanyang Technological University, Singapore. He is currently an Assistant Professor with the Department of Computer Science, City University of Hong Kong. He has authored or coauthored more than 200 refereed journal articles/conference papers. His research interests include video compression, image/video quality assessment, and image/video search and analysis. He received the Best Paper Award from IEEE VCIP 2019, ICME 2019, IEEE Multimedia 2018, and PCM 2017. His coauthored article received the Best Student Paper Award in the IEEE ICIP 2018.
\vskip2pc

\noindent\textbf{Fernanda~Andal\'{o}} is a researcher associated with the Artificial Intelligence Lab. (Recod.ai) at the Institute of Computing, University of Campinas, Brazil. Andaló received a Ph.D. in Computer Science from the same university in 2012, during which she was a research fellow at Brown University. She worked for Samsung as a researcher and was a postdoctoral researcher in collaboration with Motorola, from 2014 to 2018. Currently, she works at The LEGO Group devising machine learning solutions for digital products. She was the 2016-2017 Chair of the IEEE Women in Engineering South Brazil Section, and is an elected member of the IEEE Information Forensics and Security Technical Committee. Her research interests include machine learning and computer vision.

\vskip2pc

\noindent\textbf{S\'{e}bastien~Marcel} (IEEE Senior member) is a senior researcher at the Idiap Research Institute (Switzerland), he heads the Biometrics Security and Privacy group and conducts research on face recognition, speaker recognition, vein recognition, attack detection (presentation attacks, morphing attacks, deepfakes) and template protection. He is also Professor at the University de Lausanne (UNIL) at the School of Criminal Justice and lecturer at the Ecole Polytechnique Fédérale de Lausanne (EPFL). He received his Ph.D. degree in signal processing from Université de Rennes I in France (2000) at CNET, the research center of France Telecom (now Orange Labs). He is also the Director of the Swiss Center for Biometrics Research and Testing, which conducts certifications of biometric products. He is Associate Editor of IEEE Transactions on Biometrics and Identity Science. He was Associate Editor of IEEE Signal Processing Letters, Associate Editor of IEEE Transactions on Information Forensics and Security, a Guest Editor of the IEEE Transactions on Information Forensics and Security Special Issue on “Biometric Spoofing and Countermeasures”, and Co-editor of the IEEE Signal Processing Magazine Special Issue on “Biometric Security and Privacy”. He is also the lead Editor of the Springer Handbook of Biometrics Anti-Spoofing (Editions 1, 2 and 3).

\vskip2pc

\noindent\textbf{Anderson Rocha} is a full-professor for Artificial Intelligence and Digital Forensics at the Institute of Computing, University of Campinas (Unicamp), Brazil. He is the Director of the Artificial Intelligence Lab., Recod.ai, and was the Director of the Institute of Computing for the 2019-2023 term. He is an elected affiliate member of the Brazilian Academy of Sciences (ABC) and the Brazilian Academy of Forensic Sciences (ABC). He is a two-term elected member of the IEEE Information Forensics and Security Technical Committee (IFS-TC) and its chair for the 2019-2020 term. He is a Microsoft Research and a Google Research Faculty Fellow. In addition, in 2016, he has been awarded the Tan Chin Tuan (TCT) Fellowship, a recognition promoted by the Tan Chin Tuan Foundation in Singapore. Finally, he is ranked Top-2\% among the most influential scientists worldwide, according to recent studies from Research.com and Standford/PlosOne.

\end{document}

%% file: sections/intro.tex
In the last decade, there has been a growing expectation that Artificial Intelligence (AI) companies and researchers would dedicate their efforts to integrating humans into the digital realm or the so-called Metaverse. However, while technologies like Augmented Reality (AR) and Virtual Reality (VR) have been topics of discussion for a long time, it is only recently that significant technological advancements have made it possible to materialize such systems. 

This version of a \textit{synthetic reality} has been technically discussed at least since the 90s~\cite{Digest957:online}, and this was the main vision of where AI and related technologies would take us. However, what came as a less expected outcome is that these very technologies would inundate our physical world with content and creations, profoundly transforming our interactions with the virtual realm and reshaping how we engage with one another.

This more complex notion of \textit{synthetic reality} has been a topic of discussion by the greatest minds of our time~\cite{Titansof39:online}. On the one hand, some hold a positive perspective, recognizing the immense advantages it can bring in domains such as automation, healthcare, and innovation. On the other, a group expresses concerns about the potential perils posed by AI, such as the generation of propaganda and untruth. They even advocate for temporary halts in AI experimentation within laboratories~\cite{PauseGia10:online} to allow time for legal and ethical considerations to align with the pace of progress.

We can adopt a more pragmatic perspective and carefully embrace this emerging paradigm. This entails rapidly adapting ourselves and our societies and understanding and revitalizing our scientific endeavors.
Hence, we redefine the term ``synthetic realities'' herein as any contextual digital creation or augmentation enabled by artificial intelligence methods. These techniques/models draw upon massive amounts of data leading to a new ``reality'' or narrative regardless of its intention to deceive the individual interacting with it. When the synthetic creation harms individuals, minorities, human rights, or the rule of law, it is paramount to devise forensic techniques to pinpoint such creations and separate what is real from what is synthetic. As an example, consider the creation of a fake news piece. 
Someone could fabricate a story from scratch using a chatbot, illustrate it with a synthetic image and a video, and then broadcast it to the world as if it were real via social media.

Consequently, Forensic Science has been continually adapting to these evolving circumstances.
Rooted in the foundational principle that ``every contact leaves a trace'', coined by researcher Edmond Locard~\cite{chisum2000evidence}, Forensic Science asserts that every interaction between individuals, objects, and places leaves behind a trail of evidence. While this concept was initially centered around physical traces like fingerprints, footprints, and blood, it has recently expanded to encompass digital counterparts such as photos, audio, video, and social media posts~\cite{padilha21usp}. 

The revolving question around digital evidence is: ``Are they fake or not?'', as manipulating these multimedia assets can be quickly done with simple and inexpensive tools. Moreover, credible manipulations can be used to fabricate more believable multimedia stories. Research in Multimedia Forensics has yielded important approaches to detecting altered media~\cite{ferreira19aabc}. More specifically, progress has been made in analyzing digital media (image, video, and audio), identifying manipulations~\cite{CV-20, AF-17, KH-17, DCP+17, BP-12}, tracing provenance~\cite{HG-19, BBG+17, TCC16, BSM13, CFG+08}, and establishing links with other digital evidence~\cite{BMB+19, MBB+18, LZZ+19}.

However, the emerging concept of \textit{synthetic reality} paints an even more unsettling scenario: around 90\% of the digital content will be synthetic in the upcoming years, meaning that almost all content will be generated synthetically by definition~\cite{FinalSyn61:online}. The distinction between what is genuine and what is fake takes on a new meaning. This phenomenon becomes evident in various domains, including movies~\cite{Everyday43:online}, social media~\cite{AIGenera56:online,NvidiaUs32:online}, marketing~\cite{Virtuals79:online}, and education~\cite{Schoolsl13:online}. 

Notably, companies are now exploring the adoption of AI-generated models to promote their products~\cite{Virtuals79:online} or employing AI to simulate eye contact in video conferencing software, enhancing the sense of connection during remote interactions~\cite{NvidiaUs32:online}. Schools worldwide are banning chatbots and the use of generative AI on their networks in response to concerns about students submitting unauthentic and potentially plagiarized work~\cite{Schoolsl13:online}. The examples are many when thinking about how AI is shaping our reality.

Therefore, Forensic Science has to adapt yet again to this new reality. To expose synthetic content and tell apart malicious from harmless manipulations or even positive creations, there is one new key element: context. 
Contextual information can be leveraged to understand the semantics behind media objects. This can empower fact-checking solutions that mitigate the effect of falsified news, misinformation, and false political propaganda. To that end, and following a cognitive science interpretation, we can view digital objects through three perspectives: technological artifacts, sources of information, and platforms to convey ideas.

Traditional forensic techniques thus far have primarily focused on the first perspective. Analysts examined an asset as a digital signal and aimed to detect any possible artifact related to pixel-level or physical-level inconsistencies (e.g., concerning compression, sensor noise, illumination, shadows) to establish its authenticity. The second perspective pertains to standard fact-checking procedures going beyond multimedia forensics. When treating an object as a source of information, it is essential to identify (or know) the acquisition device and to identify the time and location where it was produced as basic steps towards a fact-checking effort. The third perspective considers a digital object as a platform for conveying ideas. Determining the intentional goal of the asset leads to answering the question of why something happened. The answer to this question and the result of forensic analyses from the other two perspectives can reveal the ultimate goal of the falsified information. For example, it can help identify if there is an ongoing campaign to bias public opinion, influence the mood of a social group, or even incite a group to articulate plans for violent acts.

The latest advancements in AI have compelled forensic techniques to navigate the intricacies between these perspectives. Taking this into consideration, we focus on studying synthetic realities in different forms of media: images, videos, audio, and text. In the remainder of this paper, we discuss how synthetic media is created, considering each of these modalities, and the implications for Digital Forensics when such creations intend to harm third parties in different ways. We explore how to detect such malicious creations and pinpoint key research challenges that lie ahead. This is particularly significant due to the remarkable progress of AI generative techniques in generating realistic content and effectively concealing the typical artifacts left behind during the creation process. Each new generative method aims at creating ever-more-believable realities, thus directly colliding with Locard's principle, the cornerstone of forensics. 


%% file: sections/syn-images.tex

The proliferation of sophisticated synthetic and manipulated media has captured people's attention worldwide. For forensic researchers, this surge in synthetic reality evokes the daunting scenario reminiscent of the early days of Digital Forensics, where image editing was recognized as a powerful tool capable of altering reality~\cite{Morris2008PhotographyasWeapon}.
While traditional manual image-editing software like Photoshop and GIMP continue to improve, they are being overshadowed (or enhanced, in the particular case of Photoshop~\cite{photoshop:online}) by the emergence of powerful AI-based generative techniques. Today, it has become remarkably effortless to transform a simple concept or idea into a realistic image, with no requirement for drawing or painting skills to produce stunning, high-quality results.

Amid this rapidly evolving landscape lies generative models. Generative images have outstanding widespread applications in entertainment, as reviving legendary artists~\cite{Ruggieri2022}; healthcare, as aiding surgeons in developing new abilities~\cite{Bohr2020}; and accessible tools, as serving people with disabilities~\cite{Carole2022}. However, many other harmful uses have been reported, such as nonconsensual DeepFake porn~\cite{Sullivan2023}, misinformation generation~\cite{Xiang2023FakeHist}, and sophisticated types of scams~\cite{Justinas2022Fraud}. Scientific integrity researchers are also concerned that such technology would create fraudulent synthetic images in science~\cite{Qi2020ScientificAIFraud, Gu2022FraudSci} and the medical area~\cite{Yisroel2019CTGAN, Mangaokar2020MedicalAttack}, in particular.

Given the alarming potential for misuse of generative models in creating synthetic realities, this section delves into state-of-the-art generative models and the detection of AI-generated images, providing perspectives on the future of synthetic images.

\subsection{Image Synthesis}
    The accelerated research on generative approaches in recent years gave birth to a plethora of AI models and techniques that are developed and open-sourced to the community. They are coupled with easy-to-use environments and applications~\cite{midjourney:online, dreamstudio:online}, allowing users to freely explore and share their creations. The increased accessibility further expanded the hype in image synthesis, fostering novel use cases and commercial applications that range from \emph{outpainting} famous art pieces~\cite{liang2022nuwa} (i.e., synthetically extending the borders of an image) to designing political campaign ads~\cite{wapo_biden_add_2023}. With the increased interest in the topic from research and industry communities and the rapid development of techniques, one can safely assume that not all synthetic images are born equally. We can categorize existing approaches by how the generation task is conditioned and what family of AI models they rely on.

\begin{figure*}[t!]
    \begin{center}
        \includegraphics[width=\textwidth]{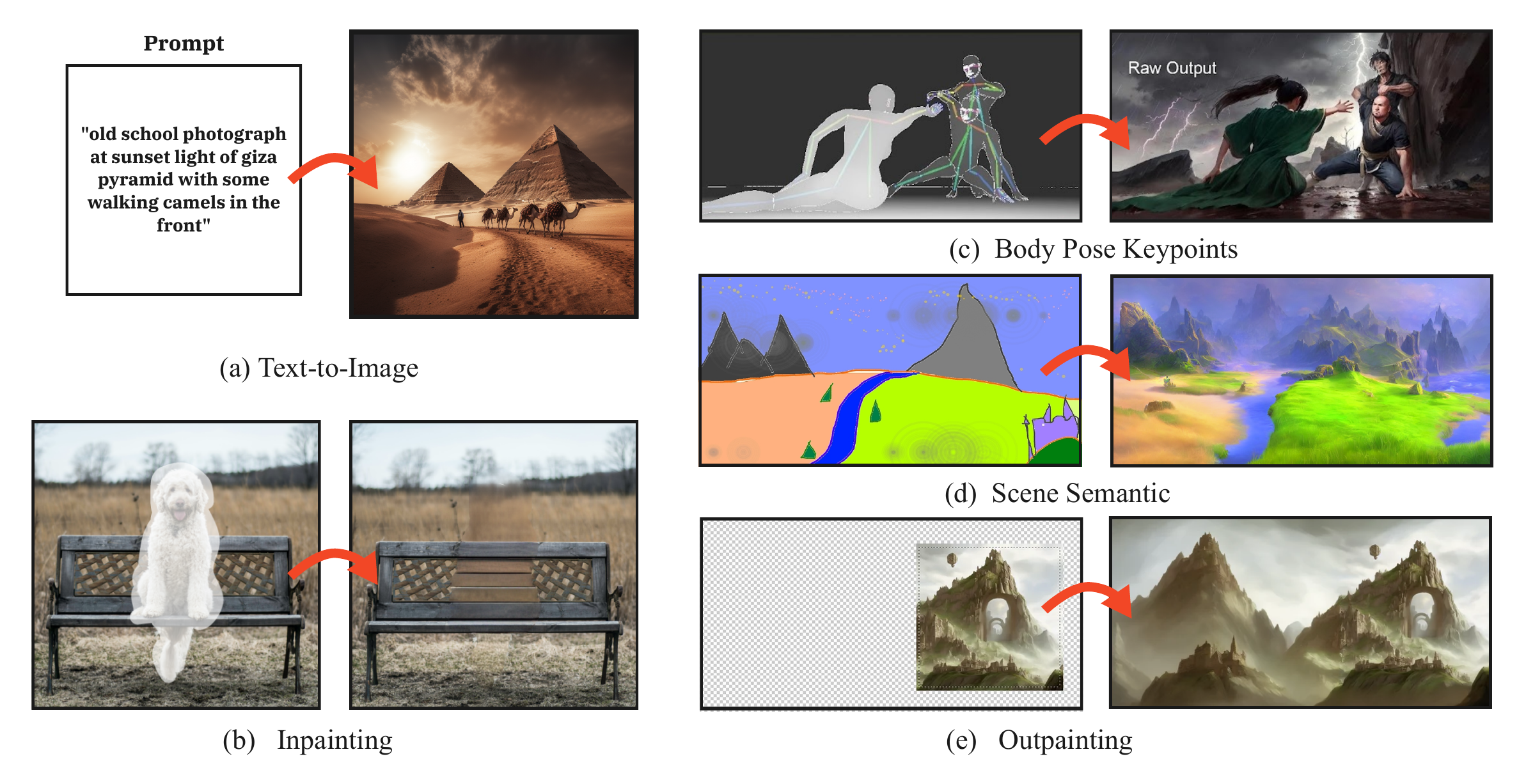}
        \caption{Examples of image synthesis conditioning. The input modality and expected output define the type of generation task performed by the model and help express desired characteristics in the synthesized creation. Examples created with and/or reproduced from~\cite{midjourney:online, automatic1111:online,stablediffusiongithub:online}.}
        \label{fig:syn-images:conditioning-tasks}
    \end{center}
\end{figure*}

    The generation task defines the goal of the method and, consequently, how it learns to map the expected input to a synthetic output image. Additionally, the input data modality conditions the generation process into expressing particular visual concepts and characteristics desired by the user~\cite{zhan2021multimodal}. The most common tasks fall into \textit{text-to-image} or \textit{image-to-image} generation. Popularized by recent applications such as MidJourney~\cite{midjourney:online} and DreamStudio~\cite{dreamstudio:online}, \textit{text-to-image} generation involves a natural language prompt describing the desired image. This often includes the object or concepts that should be created, the desired artistic style, and the feeling the composition should convey. Whereas, in \textit{image-to-image} generation, guidance may come as visual information, such as a picture, semantic map, or body pose keypoints. These may aid the generation process with information often difficult to express by natural language prompts, such as the relative positioning of the elements. We show examples of conditioning tasks and modalities in Figure~\ref{fig:syn-images:conditioning-tasks}.

    The families of AI models utilized in image generation techniques comprise several types of architectures, with the most common being Generative Adversarial Networks (GANs), Diffusion models, and Variational Autoencoders (VAE). 
    
    \textbf{GANs}~\cite{goodfellow2014generative, pan2019recent} are built on two components: a generator and a discriminator. The generator learns the underlying distribution of real examples to generate new data, while the discriminator decides whether the input is from the real data space. Through an adversarial training process, the discriminator learns to identify synthetic images, while the generator progressively improves its ability to produce high-quality images that can deceive its counterpart. Numerous GAN variants have been developed in recent years to enhance the performance and stability of image generation~\cite{radford2015unsupervised, zhang2017stackgan, karras2018progressive, zhang2022styleswin}. Among them, StyleGAN~\cite{karras2019style, karras2020analyzing, karras2021alias} allowed for intuitive control over the generated image attributes by modulating the convolutional kernels at different levels of the generator, instead of directly controlling the network input. Its successor, StyleGAN-T~\cite{stylegant2023ICML}, builds upon its architecture for text-guided image synthesis. It leverages Contrastive Language-Image Pre-Training~\cite{radford2021learning} (CLIP), a powerful text encoder that aligns textual descriptions with corresponding images. Other approaches~\cite{zhou2022towards, tao2023galip} follow a similar path, relying on CLIP to integrate natural language understanding into the image synthesis process.   

    As an alternative to the min-max optimization game of GANs, \textbf{Diffusion models}~\cite{ho2020denoising, croitoru2023diffusion} are trained to revert a stochastic diffusion process that progressively adds noise to a target image. To generate new images, the model iteratively denoises the perturbed image at each step, until a high-quality picture is reconstructed.
    By relying on a deterministic denoising function instead of  adversarial learning, 
    their training is more stable and easier to control than GANs. On the other hand, diffusion models rely on multiple network passes to reconstruct samples, constituting a 
    considerably more computationally expensive method than
    adversarial networks. To improve efficiency, Stable Diffusion~\cite{Rombach_2022_CVPR} operates on compressed latent representations instead of pixel space, mapping the denoising function to 
    smaller manifolds.
    When considering textual prompts, Imagen~\cite{saharia2022photorealistic} leverages text encoders, such as CLIP, 
    to combine them
    with multiple cascaded diffusion models to generate high-resolution outputs from text. Similarly, Ramesh et al.~\cite{ramesh2022hierarchical} train a diffusion decoder that produces images from CLIP embeddings extracted from textual prompts.  

    Another prominent family of models is \textbf{Variational Autoencoders} (VAE)~\cite{kingma2013auto}. 
    Autoencoders follow an encoder-decoder architecture that projects the input data into a low-dimensional latent space and learns to reconstruct the original input from it. VAE, in turn, extends upon autoencoders by adding a probabilistic component to the latent representation. Instead of learning a deterministic encoding for each input, the network learns the parameters of a probabilistic distribution that models the latent space. By doing so, it can sample from the latent space distribution to generate new samples. Constraining the low-dimensional latent space further, VQ-VAE~\cite{van2017neural, razavi2019generating} uses vector quantization to learn discrete latent variables, which improves the interpretability of the learned concepts and allows for easily manipulating them when generating new compositions. Building on top of the previous technique, DALL-E~\cite{ramesh2021zero} and CogView~\cite{ding2021cogview} address text-to-image generation by combining the rich representations learned by variations of VQ-VAE with the predictive capabilities of Transformers~\cite{vaswani2017attention}. Both approaches use Transformer modules to predict the best image tokens from the VQ-VAE codebook, given a textual token and previously selected visual tokens.
    This results in
    coherent and semantically meaningful synthesized creations.

    Each of these AI model families has its 
    pros and cons.
    GANs excel at generating sharp and visually compelling images, but they may suffer from mode collapse and training instability issues when used in large and diverse datasets~\cite{pan2019recent}. Diffusion models offer a powerful approach to generating high-quality visual data, but they can be computationally expensive due to their iterative nature. VAEs provide a more straightforward training process with a clear optimization objective, but they may generate less sharp images than GANs and diffusion models. Nonetheless, all of them made significant advancements in the field of image generation, enabling the synthesis of realistic and diverse pictures. As the interest in this area increases, more advances will come, and the realism gap between real and synthetic data will shorten to the point that distinguishing between them will be challenging. This poses numerous problems in assessing the reliability and authenticity of visual content in an increasingly digital world. With this in mind, we discuss existing forensic approaches that may help to identify synthetic creations in the next section. 

\subsection{Synthetic Images Detection}
\begin{figure*}[t!]
    \includegraphics[width=\linewidth]{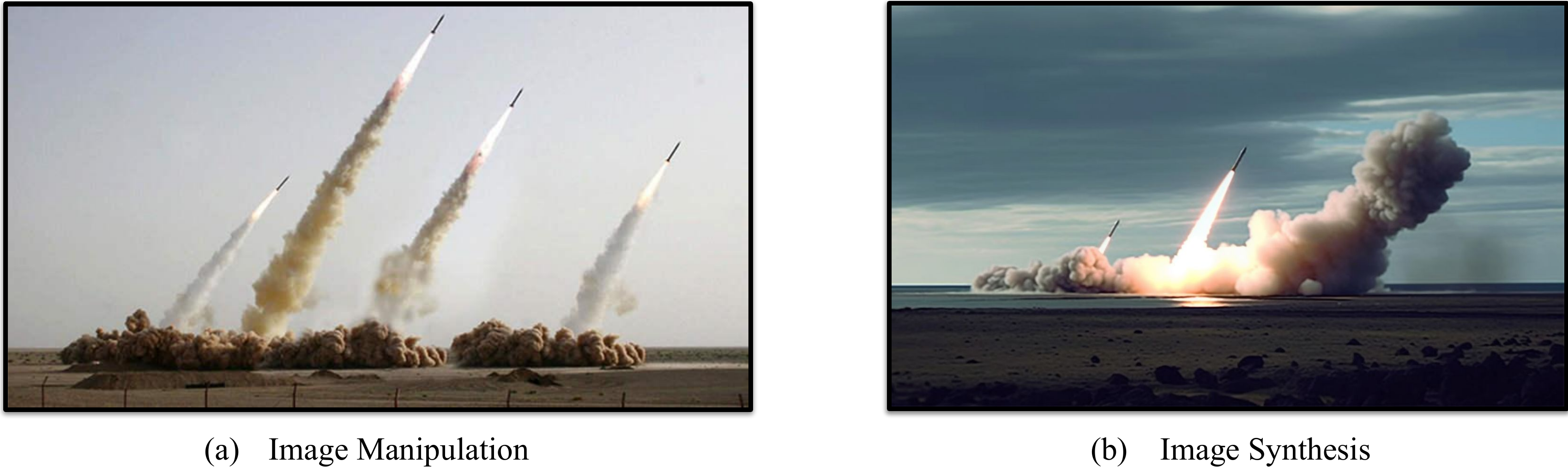}\label{fig:img_syn_old_new_a}
    \caption{Classic manual image manipulation versus modern image synthesis.
    The image on the left (a) represents a well-known case of state-level manual image manipulation, which was misleadingly published by multiple news websites as genuine in 2008.
    In contrast, we generated the image on the right (b) with MidJourney by using a prompt as simple as ``missile test''.
    Cases like the former one currently represent an even greater challenge to authenticity verification.
    }
    \label{fig:img_syn_old_new}
\end{figure*}

In contrast to old-fashioned types of image manipulation, modern synthetic images take their realism to higher standards. Figure~\ref{fig:img_syn_old_new} compares classic Digital Forensics cases with those created by generative models, showcasing the remarkable advancement achieved. The level of refinement and potential harm associated with these synthetic images raise concerns about the capability of Digital Forensics to identify such content. However, we anticipate that \textbf{Locard's exchange principle still holds for synthetic imagery}, with forensic traces taking the form of visual inconsistencies and artifacts left by the generation process. Nevertheless, as these models continue to evolve and new forms of counter-forensics attacks emerge, the validity of this claim may be challenged.

In this section, our analysis focuses on examining possible traces left by GANs and other generative models. We categorize our study based on the types of forensic evidence utilized for detection, namely \textbf{Visual Artifacts} and \textbf{Noise Fingerprints}.

\subsubsection{Visual Artifacts}
\begin{figure*}[t!]
    \centering
    \includegraphics[width=\textwidth]{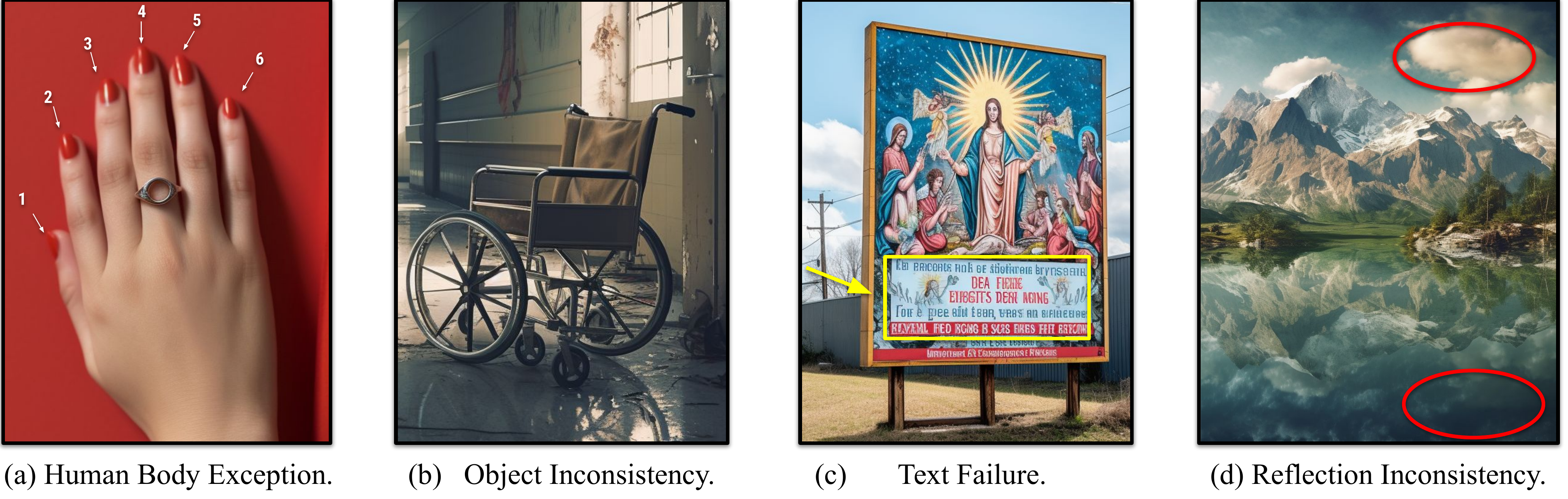}
    \caption{Examples of generated image inconsistencies.
    All images were generated with version 5.1 of the Midjourney model -- the latest one released 
    at the writing of this article. In (a), an unnatural synthetic hand with six fingers. In (b), a synthetic wheelchair with inconsistent design; 
    the seat orientation does not match the wheels' position.
    In (c), a synthetic 
    billboard with text that makes no sense and presents aberrant letters.
    In (d), a synthetic paisage with cloudy skies and mountains by a lake; the highlighted cloud is not congruently reflected on the lake's surface. To generate these images, we used the following prompts: (a) ``lady's hand with a ring on it'', (b) ``wheelchair in a hospital'', (c) ``outdoor sign with a religious statement on it'', and (d) ``realistic photo; mountains with a lake at the bottom''.
    }
    \label{fig:syn-img-failures}
\end{figure*}

Despite the impressive realism of cutting-edge synthetic images, a closer look reveals various aberrational results and visual inconsistencies. Borji~\cite{Borji2023FailuesImageDeepFake} has presented several image failures that can occur when generating synthetic content, even with 
recent
generative models such as DALL-E 2~\cite{ramesh2022hierarchical}, Midjourney~\cite{midjourney:online}, and StableDiffusion~\cite{stablediffusiongithub:online}. These failures 
may occur
in the background, reflections, lighting, shadows, text, body parts, and objects, as depicted by Figure~\ref{fig:syn-img-failures}.
This figure illustrates clues that have been explored by Digital Forensics researchers to detect synthetic content.

Farid, for instance, analyzed the 3D illumination~\cite{Farid2022Lighting} and geometric~\cite{Farid2022Geometry} consistency of structures and objects in a photograph generated by state-of-the-art generative models. His analysis employed classic digital forensics techniques for on-scene illumination and 3D geometric analysis, similar to those used in previous classic forensic works~\cite{Bertamini2003, Ostrovsky2005, Carvalho2013Illumination}. By doing so, Farid showed that while the local structures in the photo may be globally consistent, they exhibit local inconsistencies that serve as valuable clues for forensic analysts.

However, as generative models continue to evolve, it is expected that the visual inconsistencies and artifacts observed in synthetic images will eventually become rarer or imperceptible, as demonstrated in the case of the AI-synthetic faces examined by Nightingale and Farid~\cite{Nightingale2022}. This situation asks for other strategies that rely on other types of forensic clues, such as the noise left by the generation processes.

\subsubsection{Noise Fingerprint}
As image synthesis is rapidly improving, it is crucial to employ a variety of alternative detectors that explicitly exploit different characteristics of synthetic images. Therefore, as an alternative to visual artifacts, noise-based detectors have been a promising path to expose synthetic content.

In this direction, Marra et al.~\cite{Marra2019GanFingerprints} investigated statistics-based techniques to detect potential noise fingerprints left by GANs on their generated content. By utilizing photo response non-uniformity (PRNU) analysis, similar to camera attribution methods, they discovered a correlation between residual noise patterns and specific GAN models. Furthermore, Marra et al.~demonstrated the feasibility of differentiating between distinct GAN models used for image synthesis through residual noise analysis, enabling GAN model attribution.

In a similar study, Mandelli et al.~\cite{Mandelli2022} revealed that comparable residual noise patterns could be leveraged to identify GAN-generated scientific images, indicating the potential extension of this approach to other image types beyond natural images. These findings emphasize the applicability of residual noise analysis in detecting and identifying synthetic images, contributing to the field of Digital Forensics. Noise signatures allied with visual clues were explored in tandem by Kong et al.~\cite{kong2022detect}, showing that combining different evidence might be the way forward in dealing with the challenges of synthetic realities detection.   

In a more recent investigation, Corvi et al.~\cite{corvi2023intriguing} examined the presence of fingerprints left by state-of-the-art generative models, including GAN-based and Diffusion-based models. Their findings reveal that \textbf{no generative model appears to be completely artifact-free} at present. Both GAN-generated and Diffusion-generated images exhibit anomalous periodic patterns in the Fourier spatial domain. 

However, as highlighted by Gragnaniello et al.~\cite{Gragnaniello2021}, such artifacts may be challenging to detect when post-processing operations are applied, such as image resizing and compression. These operations are frequently employed on social media platforms to save storage and speed up sharing, further complicating the identification of these visual irregularities.
Besides investigating how post-processing operations impact synthetic image detectors, researchers have also identified the potential for synthetic image detectors to be deceived through counter-forensics attacks.

\subsubsection{Counter Forensics}
In a recent study that challenges noise-based detection methods, Osakabe et al.~\cite{Osakabe2021} developed a GAN model that can generate images without ``checkerboard artifacts'', a specific type of artifact in the Fourier domain that is common in synthetic images. They achieved this by incorporating a fixed convolutional layer into every upsampling and downsampling layer of the GAN architecture. Remarkably, their model successfully fooled a detector that previously identified fake images with 92\% accuracy, reducing the accuracy to a mere 12\%.

Similarly, Cozzolino et al.~\cite{Cozzolino2019SpoCSC} demonstrated that synthetic image detectors could be fooled by 
transferring 
residual noise fingerprints from real cameras onto GAN-generated images. This process produces a spoofed image that can avoid 
accurate GAN detectors and camera-model identifiers, causing the image to be misattributed as originating from the 
transferred camera model.

As counter-forensic attacks indicate, synthetic image detection presents multiple research challenges to ensure media integrity and prevent images from being used by a malicious actor.

\subsection{Challenges and Directions}



As with classic edited images, creating synthetic realities using cutting-edge generative models has sparked ethical debates and raised concerns about their use. Once again, Digital Forensics plays a crucial role in this debate by investigating ways to detect the traces of artificial intelligence techniques left in these images. While most synthetic images can be identified through a close look into 
scene inconsistencies and object aberrations, as depicted in Figure~\ref{fig:syn-img-failures}, the advancement of image synthesis will inevitably render these visual incongruences invisible to the naked eye.

Consequently, researchers also rely on new types of fingerprints inherently left by the generation process, such as specific patterns on the Fourier spatial spectrum and residual noise analysis. Some of these artifacts can be compared to the PRNU noise left by camera sensors. They can aid forensic analysis not only in detecting fake images but also in identifying the specific generative model used to render them (source attribution). However, traces alone provide a vulnerable target for synthetic image detection, as they can easily be manipulated to deceive accurate fake image detectors. Such attacks may involve common post-processing operations or sophisticated techniques like camera noise 
transference. Therefore, it is imperative for forensic researchers to develop robust techniques capable of detecting and distinguishing these artifacts, even in the presence of common post-processing operations or more sophisticated attacks.

A more challenging and socially responsible aspect of forensics involves preventing the harmful applications of synthetic images. Given the ease with which such content can be created and shared, it is essential for forensics researchers to design traceable techniques 
whose synthetic images can be readily distinguished. Traceable evidence would assist analysts in swiftly identifying the source and author of such content, thereby preventing its widespread dissemination. In this vein, researchers have developed deep learning-based watermarking approaches~\cite{luo2020distortion, ahmadi2020redmark, fernandez2023stable} to identify synthesized content. These methods use encoder layers to imbue watermark information in the image pixels without perceptually altering its content. For successfully marking creations, researchers aim to be robust to most online alterations, such as compression, cropping, and intensity changes. Unfortunately, most approaches act from the generator's perspective, either adapting existing models or adding external components in the generation process to enable watermarking. This might be viable for well-established and commercial applications (e.g., Midjourney~\cite{midjourney:online}) but will be hardly used with open-sourced models that are trained and distributed by the community.



%% file: sections/syn-videos.tex
The ability to generate realistic and 
useful videos holds immense value across various application domains such as entertainment, virtual reality, and education~\cite{ranzato2014video, vondrick2016generating}. Undoubtedly, video generation techniques have made significant positive contributions in these domains. The advancements have opened up new avenues for creativity, synthetic realities, and immersive experiences. However, it is essential to acknowledge that along with their benefits, these techniques also raise potential security concerns. Synthesize realistic videos can be exploited for malicious purposes, such as financial fraud and the dissemination of fake news. Consequently, ensuring the integrity and authenticity of digital content becomes increasingly critical.

Generally speaking, video synthesis can be divided into video generation and text-to-video synthesis. Previous methods for video generation mainly employ GANs~\cite{vondrick2016generating, tulyakov2018mocogan, tian2021good} and VAEs~\cite{yan2021videogpt, le2021ccvs} to generate videos. But with the advent of diffusion models, recent methods explore them to generate more realistic videos~\cite{blattmann2023align, voleti2022masked, yang2022diffusion, ramesh2022hierarchical, singer2022make}. On the other hand, text-to-video synthesis incorporates text information to guide the model in generating video content that is responsive to specific demands. Analogously, previous text-to-video synthesis methods have resorted to GANs~\cite{li2018video, pan2017create} and diffusion models \cite{ho2022imagen, an2023latent, esser2023structure, he2022latent, khachatryan2023text2video}, achieving exceptional generation quality in terms of fidelity, resolution, and temporal consistency. 

In this section, we provide a review of the methods for video generation, text-to-video synthesis, synthetic video detection and discuss their challenges. We further outline possible future research directions for synthetic video generation and detection techniques.  

\subsection{Video Generation}

In the pursuit of advancing video synthesis, previous research has extensively explored diverse generative models, including Generative Adversarial Networks (GANs)~\cite{saito2017temporal, vondrick2016generating}, autoregressive models~\cite{srivastava2015unsupervised, yan2021videogpt}, and implicit neural representations~\cite{skorokhodov2022stylegan, yu2022generating}. However, recent attention has been drawn to the exceptional achievements of diffusion models in visual data synthesis. Several notable works propose outstanding video generation methods and investigate their practical applications~\cite{blattmann2023align, hoppe2022diffusion, voleti2022masked, wu2022tune}. For instance, a pioneering work on diffusion video generation~\cite{ho2022video} primarily focuses on network architecture modifications to extend image synthesis to video. The 3D U-Net is adopted~\cite{cciccek20163d} and achieves outstanding generation results in two cases, including unconditional and text-conditional video generation. For longer video generation, they apply an autoregressive approach, where subsequent video segments are conditioned on the preceding ones. Another example of a diffusion video generation work~\cite{yang2022diffusion} adopts frame-by-frame video generation models. To evaluate different prediction strategies, the authors conduct an ablation study to determine whether predicting the residual of the next frame yields superior results compared to predicting the actual frame. 

Furthermore, Hoppe $et~al.$~\cite{hoppe2022diffusion} introduced the Random Mask Video Diffusion (RaMViD) technique, which can be utilized for both video generation and infilling tasks. The unmasked frames are used to enforce conditions on the diffusion process, while the masked frames undergo diffusion through the forward process. By employing this training strategy, RaMViD demonstrates outstanding video generation quality. These recent advancements in diffusion-based video generation highlight the potential of this family of models to push the boundaries of video synthesis, addressing the challenges of generating realistic and diverse video content in synthetic realities.

\subsection{Text-to-video Synthesis}
Text-to-video models are highly data-hungry, which require massive amounts of data to learn caption relatedness, frame photorealism, and temporal dynamics~\cite{ge2023preserve}. However, video data resources are comparatively more limited in terms of style, volume, and quality. This scarcity of video data poses significant challenges for training text-to-video generation models. To overcome these challenges, additional controls are often incorporated to enhance the responsiveness of generated videos to user demands~\cite{mathieu2015deep, pan2017create, wang2018video}. 

Early text-to-video generation models heavily relied on convolutional GAN models combined with Recurrent Neural Networks (RNNs) to capture temporal dynamics~\cite{li2018video, pan2017create}. Despite the introduction of complex architectures and auxiliary losses, GAN-based models exhibit limitations in generating videos beyond simplistic scenes involving digit movements or close-up actions. To that end, recent advancements in the field have aimed to extend text-to-video generation to more diverse domains using large-scale transformers~\cite{yu2023magvit} or diffusion models~\cite{ho2022imagen}. These approaches provide promising directions for generating more complex and realistic video content by leveraging the expressive power of these advanced network architectures.

However, modeling high-dimensional videos and addressing the scarcity of text-video datasets present considerable challenges in training text-to-video generation models from scratch. To tackle this issue, most approaches adopt a transfer learning paradigm, leveraging pre-trained text-to-image models to acquire knowledge and improve performance. For instance, CogVideo~\cite{hong2022cogvideo} builds upon the pre-trained text-to-image model CogView2~\cite{ding2022cogview2}, while Imagen Video~\cite{ho2022imagen} and Phenaki~\cite{villegas2022phenaki} employ joint image-video training techniques to leverage pre-existing visual representations. In contrast, Make-A-Video~\cite{singer2022make} focuses on learning motion solely from video data, reducing the reliance on text-video pairs for training.

Another key consideration in video synthesis is the high computational cost associated with generating high-quality videos. To mitigate this issue, latent diffusion has emerged as a popular technique for video generation, as it offers a computationally efficient alternative~\cite{an2023latent, blattmann2023align, esser2023structure, he2022latent}. Various powerful but computational-efficient methods, such as MagicVideo~\cite{zhou2022magicvideo}, which introduces a simple adaptor after the 2D convolution layer, and Latent-Shift~\cite{an2023latent}, which incorporates a parameter-free temporal shift module, have successfully utilized latent diffusion for video synthesis. Additionally, PDVM~\cite{yu2023video} adopts a novel approach of projecting the 3D video latent space into three 2D image-like latent spaces, further optimizing the computational cost of the video generation process.

Despite the active research in text-to-video generation, existing studies have predominantly overlooked the interplay and intrinsic correlation between spatial and temporal modules. These modules play crucial roles in understanding the complex dynamics of videos and ensuring coherent and realistic video generation. 

\subsection{Synthetic Video Detection}


Again, detecting synthetic videos relies on the fingerprints left by generative models, which have been explored extensively in the context of synthetic image detection. Existing research in synthetic image detection has shown promise by identifying inconsistencies in illumination and geometric structure~\cite{Farid2022Lighting, Farid2022Geometry}, as well as specific noise patterns in the Fourier domain~\cite{corvi2023intriguing}. However, the extension of image-based detection techniques to videos is still in its early stages.

One straightforward approach for extending image-based detection to videos is through frame-level voting, where each frame is individually analyzed and classified as real or synthetic. However, exploiting temporal information, such as temporal coherence, presents a significant challenge. The temporal domain contains valuable cues that can aid in distinguishing synthetic videos from real ones. For instance, temporal coherence refers to the consistent motion and flow of objects across frames in a real video. Detecting such temporal fingerprints could provide valuable insights into the authenticity of a video. The  temporal coherence, which is a significant challenge in the field of video synthesis \cite{wang2023videofactory}, shall also be vital for synthetic video detection. Consequently, exploiting the temporal inconsistency can be utilized to identify generated videos. Currently, there is no existing method specifically designed for detecting video synthesis. However, there have been relevant work on detecting Deepfake videos using generic architectures. For instance, 3DCNN \cite{zhang2021detecting}, RNN~\cite{amerini2020exploiting}, LSTM~\cite{chintha2020recurrent}, and temporal transformer~\cite{mittal2020emotions} have been widely employed for video-level Deepfake detection. Additionally, some studies focus on detecting deepfake videos by examining specific temporal artifacts like lip movement~\cite{haliassos2021lips}, rPPG artifacts~\cite{qi2020deeprhythm}, and head pose inconsistency~\cite{yang2019exposing}. Thus, how to explore the temporal artifacts ($e.g.$, unnatural motions) in generated videos to detect video synthesis is an intriguing task.

This temporal aspect of video analysis introduces additional complexities compared to image analysis. Generated video content can exhibit various artifacts due to the processing techniques employed, including both handcrafted designs and deep neural networks. These artifacts, including blur, compression artifacts, and noise, can be intentionally or unintentionally injected during the video generation process. Consequently, these artifacts may pose significant obstacles to the detection of synthetic videos, particularly when the detection model is trained solely on high-quality video data. 

To address this challenge, detection models need to be highly generalized to handle data from different domains. The models must be capable of recognizing and adapting to various levels of quality, distortion types, and content sources. This requirement calls for the application of domain generalization techniques, which enable the model to generalize well beyond the training data distribution. By training the model on a diverse range of video data, encompassing different quality levels, distortion types, and content sources, the detection system can become more robust and effective in identifying synthetic videos across a wide range of scenarios.

Overall, the extension of image-based detection techniques to videos presents a meaningful yet challenging direction for research. Leveraging temporal fingerprints and addressing the presence of artifacts in generated videos require novel approaches and further exploration. Developing detection models that can effectively analyze and distinguish synthetic videos while being adaptable to various domains will play a crucial role in combating the increasing threat of synthetic videos in today's digital landscape.


\subsection{Challenges and Directions}
Recent advancements in diffusion models have revolutionized text-to-video synthesis, achieving remarkable capabilities that surpass previous state-of-the-art approaches and deliver unprecedented generative performance. This breakthrough has significantly enhanced the quality and fidelity of generated videos. However, as we delve deeper into this domain, it becomes evident that there is a need for further research and development. 

Despite the demonstrated success in generated image content in the past few years, video generation is still in its infancy. As we have discussed, the challenges of video generation mainly lie in the following three aspects: (1) lacking large-scale, diverse, and in-the-wild video datasets; (2) demanding computational costs; and (3) unstable in synthesizing  coherent content from both spatial and temporal perspectives. 

In the realm of synthetic video detection, existing methods predominantly focus on identifying face forgery, where the manipulation targets primarily involve facial features and expressions. However, to address the evolving landscape of synthetic videos and their potential threats, it is imperative to explore detection techniques that encompass a broader range of scenes and contexts. Detecting synthetic videos with diverse scenes, objects, and backgrounds poses an interesting avenue for future advancements in forensic research. By expanding the scope of detection techniques, we can develop robust and comprehensive methods that effectively identify and mitigate the risks associated with the increasing sophistication of synthetic videos.

%% file: sections/syn-audio.tex
Synthetic realities in audio are an emerging technology transforming how we experience sound. From augmented and virtual reality to interactive audio installations, synthetic realities offer a new dimension to our auditory senses. These immersive audio experiences create a simulated environment that can transport listeners to different worlds, trigger emotions, and enhance storytelling. With the advancements in audio technology and the increasing demand for immersive experiences, synthetic realities are poised to revolutionize the entertainment, gaming, and education industries. While the rise of synthetic audio technology has brought about significant benefits to various fields, it also presents a considerable threat to the integrity of our society. One of the most concerning implications is the potential misuse of this technology by malicious actors, who can exploit it for nefarious purposes such as telecommunication fraud. Cai \textit{et al.} \cite{cai2021generative} have highlighted the dangers of using generative models to create fake audio that can deceive individuals and organizations, leading to financial losses and reputational damage. The use of synthetic audio in such fraudulent activities underscores the urgent need for developing robust and reliable methods for detecting and mitigating the harms of this technology.  In this section, we propose to survey synthetic realities in audio and dive into the possibilities and challenges of this rapidly evolving field.

\subsection{Synthetic Audio Generation}

Audio synthesis is a vital and rapidly evolving research area with a wide range of applications, including text-to-speech (TTS), speech enhancement, voice conversion, and binaural audio synthesis. In the field of TTS, previous works have extensively utilized deep learning-based architectures such as WaveNet~\cite{oord2016wavenet} and Clarinet~\cite{ping2018clarinet}, as well as transformer models like FastSpeech~\cite{ren2019fastspeech} and Neural TTS~\cite{li2019neural}, and variational autoencoder (VAE) approaches such as MultiSpeech~\cite{guo2022multi} and Hierarchical VAE~\cite{hsu2018hierarchical}. Recently, diffusion models have gained prominence in addressing TTS problems, with notable contributions from WaveGrad~\cite{chen2020wavegrad}, DiffWave~\cite{kong2020diffwave}, Gradient Flow~\cite{popov2021grad}, and Diffusion TTS~\cite{jeong2021diff}. 

Speech enhancement techniques aim to improve speech recognition system performance by mitigating the impacts of ambient noise. The advancement of generative models has led to the development of various approaches for speech enhancement. These include GAN-based methods like MetricGAN~\cite{fu2019metricgan} and Speech Enhancement GAN~\cite{lin2019speech}, as well as diffusion-based models such as Storm~\cite{lemercier2022storm}, Conditional Diffusion~\cite{lu2022conditional}, and Cold Filter~\cite{yen2023cold}. These models have exhibited general and robust speech enhancement performance.

Voice conversion, another critical task in speech synthesis, aims to transform the voice of one speaker into that of another. Different approaches have been explored for voice conversion, including transformer-based models (VoiceFilter~\cite{huang2019voice}), GAN-based methods (CycleGAN-VC~\cite{kaneko2019cyclegan}, VoiceGAN~\cite{hsu2017voice}), and VAE-based techniques (ACVAE~\cite{kameoka2019acvae} and Neural Voice Cloning~\cite{choi2021neural}). These methods facilitate the manipulation of speaker characteristics while maintaining the linguistic content of the speech.

Lastly, binaural audio synthesis~\cite{richard2021neural, leng2022binauralgrad} focuses on transforming mono audio signals into binaural audio, which enables accurate sound localization and immersive auditory experiences. By simulating the perception of sound through two ears, binaural audio synthesis contributes to creating a more realistic and interactive auditory environment.

Overall, the continuous advancements in deep learning, generative models, and various synthesis techniques have significantly expanded the possibilities and applications of audio synthesis, enhancing the quality, naturalness, and versatility of synthesized speech and audio.

\subsection{Synthetic Audio Detection}



Existing methods for synthetic audio detection can be categorized into two different streams: feature-based and image-based methods~\cite{bhagtani2022overview}.
Feature-based approaches describe the audio through signal features such as Mel frequency cepstral coefficient (MFCC) and constant Q cepstral coefficient (CQCC) \cite{yang2018extended,todisco2016new}.
These features are then fed into typical classifiers (e.g., support vector machines)~\cite{wu2015asvspoof} and deep neural networks~\cite{mehrish2023review}), which are trained to detect synthetic audio.
Image-based methods, on the other hand, utilize either spectrogram images \cite{fathan2022mel,lim2022detecting} computed from the audio signal and use them as the input for deep neural networks to extract discriminative information for synthetic audio detection. The aforementioned techniques have also been widely applied to deepfake detection related to synthetic audio \cite{chen2020generalization} (more detail can be found in Section VII).



\subsection{Challenges and Directions}
Despite the progress made in synthetic audio detection, there are still challenges to overcome. One of the critical challenges is the availability of large datasets of synthetic audio that can be used to train detection models effectively \cite{tan2021survey}. Generating a large dataset of synthetic audio can be time-consuming and resource-intensive. Additionally, the increasing complexity and sophistication of synthetic audio algorithms may require more advanced detection methods that can keep up with these advancements.

While there are challenges to overcome, such as the availability of large datasets and the need for more advanced detection methods, recent research has shown promising results in developing more effective detection techniques. As synthetic audio technology advances, it is essential to continue developing and improving detection methods to prevent the misuse of deepfake audio and ensure that this technology is used safely and responsibly.



%% file: sections/syn-text.tex
Large Language Models (LLMs) have revolutionized artificial intelligence, marking a significant milestone in the field. Since the rise of GPT models~\cite{radford2019language,brown2020language}, competitors from other companies like Google and Microsoft have also developed their own LLM, including Gopher~\cite{rae2021scaling}, GLaM~\cite{du2022glam}, LaMDA~\cite{thoppilan2022lamda}, Megatron-Turing NLG~\cite{smith2022using}, Chinchilla~\cite{hoffmann2022training}, PaLM~\cite{chowdhery2022palm}, BloombergGPT~\cite{wu2023bloomberggpt}. Some open-source models have also been released, including Jurassic-1~\cite{lieber2021jurassic}, GPT-J~\cite{wang2021gpt}, GPT-NeoX~\cite{black2022gpt}, BLOOM~\cite{scao2022bloom}, OPT~\cite{zhang2022opt}, LLaMA~\cite{touvron2023llama}, and OpenAssistant~\cite{kopf2023openassistant}. Although there are many variants of LLMs models, most of them have the same architecture of the GPT-2 model~\cite{radford2019language}, which is an auto-regressive model (decoder-only transformer) that generates the next word based on previous inputs.

\begin{figure*}[t!]
\centering
   \subfloat[Poem ``Thoughts'' - Human Version.]{
    \fbox{\parbox{0.4\linewidth}{\footnotesize{
What kind of thoughts now, do you carry\\
   In your travels day by day\\
Are they bright and lofty visions,\\
   Or neglected, gone astray?\\

Matters not how great in fancy,\\ 
    Or what deeds of skill you've wrought;\\
Man, though high may be his station,\\
    Is no better than his thoughts.\\

Catch your thoughts and hold them tightly,\\
   Let each one an honor be;\\
Purge them, scourge them, burnish brightly,\\
   Then in love set each one free.\\
}}}}
    \label{fig:poems_human}
  \hfill
    \subfloat[Poem ``Thoughts'' -  LLM Version.]{
     \fbox{\parbox{0.4\linewidth}{\footnotesize{
     In the realm of silent contemplation,\\
    Thoughts weave their intricate creation.\\
    Whispers of ideas, ethereal and vast,\\
    A symphony of musings from the past.\\
    
    They flutter like butterflies in the mind,\\
    Seeking connections, truths to find.\\
    Embrace the thoughts that come your way,\\
    For within their essence, you may find,\\
    The power to shape your world, your mind.\\
    
    In the realm of thoughts, oh, so grand,\\
    Unleash your creativity, take a stand.\\
    Let them flow, let them soar,\\
    And in their symphony, discover more.
    }}}}
    \label{fig:poems_machine}
  \caption{Human versus Machine poem writing. (a) is a public-domain poem entitled Thoughts from Myra Viola Wilds; (b) is a poem written by chatGPT when asked to write a poem entitled `Thoughts'.}
  \label{fig:poems}
\end{figure*}

Figure~\ref{fig:poems} illustrates how these models have reached a quality comparable to humans, even in highly complex tasks such as poem writing, as both poems seem to be written by an excellent poet. The widespread adoption of LLMs has been observed across diverse domains, including Medicine~\cite{ChatGPTW64:online}, Journalism~\cite{Manjoo_2023}, and Science~\cite{Tran_2023}. These powerful tools possess immense potential to enhance human capabilities in various areas, ranging from code development (e.g., GitHub Copilot\footnote{\url{https://github.com/features/copilot}}) to combating online hate speech and harassment (e.g., Cohere Classify \footnote{\url{https://txt.cohere.com/content-moderation-classify/}}).

While they offer numerous benefits, there is also a risk of them being used to produce harmful content, either deliberately by malicious actors or inadvertently due to their inherent flaws. In this section, we will investigate potential flaws and damaging applications of LLMs through a forensic lens. Our exploration will encompass the emergence of threats, machine-generated text detectors, and the underlying research challenges.

\subsection{Large Language Models Threats}
Text generation, like any form of machine-generated content, possesses inherent scalability, granting it the power to be employed in both beneficial and detrimental ways. While AI-generated text may still exhibit semantic flaws or hallucinations~\cite{Alkaissi2023}, it has become increasingly difficult to differentiate between human and machine-generated text. This convergence of quality between human and AI-generated text poses a significant concern, particularly in the hands of malicious actors.

In a comprehensive study about computer-generated text threat modeling, Crothers et al. ~\cite{Crothers2022MachineGT} categorize various types of attacks facilitated by large language models (LLMs). They group these attacks into four primary threats: (1) Facilitating Malware and Social Engineering; (2) Spam and Harassment; (3) Online Influence Campaigns; and (4) Exploiting AI authorship. While we will enumerate some of these threats within this section, it is worth noticing that LLMs have opened up a wide range of possibilities for malicious actors, extending beyond the scope of our enumerated list.

\subsubsection{Facilitating Malware and Social Engineering}
This threat makes use of LLMs for facilitating scalable and customizable scams, making them a significant threat in the realm of malware and social engineering~\cite{Giaretta2020ComunnityTP}. By leveraging techniques like fine-tuning and prompt engineering, malicious actors can generate tailored scams that are highly convincing and appealing to specific targets or communities. For instance, by incorporating social media data from a target's profile, such as their interests, lifestyle, and social connections, LLMs can create more sophisticated and personalized scams that manipulate individuals into taking harmful actions or providing sensitive information. 

Another notable threat within this category is \textit{Data Poisoning}. It involves the injection of exploitable data into the training process of LLMs or fine-tuning them with malicious intent. Schuster et al.~\cite{Schuster2021AutocompleMe} demonstrated a possible attack that poisons code-completion models (e.g., GitHub Co-pilot), making them include code vulnerabilities in their output, which attackers can later exploit. Such attacks open an important discussion about training datasets, as they are often gathered from the web without any rigorous curation.

\subsubsection{Spam and Harassment} 
This threat weaponizes LLMs through trolls and hateful communities to propagate toxic content, disseminate misinformation, and target specific communities for harassment. An example of such an attack is the creation of GPT-4chan, as highlighted by Yannic Kilcher in his video ``This is the worst AI ever''~\cite{Kilcher2022}. Kilcher fine-tuned GPT-J using data collected from the /pol/ channel on 4chan, a controversial online platform forum channel. The resulting model was used to interact with users on the same channel, encapsulating the offensive, nihilistic, and trolling nature that characterizes many /pol/ posts~\cite{Kilcher2022}. Kilcher's experiment raised the alarm to the scientific community on the ease with which LLMs can be misused and the potential consequences of such actions~\cite{Kurenkov_2022}. It emphasized the need for careful consideration and ethical responsibility when deploying and sharing LLMs, as they can be harnessed to amplify harm and propagate hateful ideologies.

\subsubsection{Online Influence Campaigns}
The utilization of LLMs for spreading fake news and manipulating public opinion has emerged as a significant concern. Political campaigns, in particular, could be a perilous case through the use of LLMs. Bai et al.~\cite{Bai2023} have demonstrated the susceptibility of individuals to persuasion on political matters when exposed to tailored messages generated by LLMs. Malicious actors could use this phenomenon in a devastating scenario to influence elections and other democratic processes.

\subsubsection{Exploiting AI authorship}
A intriguing threat of LLMs is academic articles generations. One can remind the case of SCIgen (2005), where MIT graduate students developed a system for automatically generating computer science papers, demonstrating the vulnerability of academic conferences to such submissions~\cite{Scigen_2005}. There is a growing concern that LLMs could be exploited to generate much more sophisticated fake articles than SCIGen, compromising scientific integrity. Research integrity experts fear that paper mills\footnote{Potentially illegal organizations that offer ghostwritten fraudulent or fabricated manuscripts~\cite{Byrne2020}.} will improve their production in quality and quantity by using LLMs~\cite{Tran_2023}.

\subsection{Detection Methods}
A few detection methods have been proposed for LLMs generated content. One of the pioneering approaches is GROVER, proposed by Zellers et al.~\cite{zellers2019defending}. GROVER was capable of generating fluent and highly realistic fake articles using LLMs, which motivated the authors to explore detection techniques for such content.
Zellers et al. found that employing the same model used for generating the text achieved higher detection accuracy compared to using a different one. Their results demonstrated an impressive accuracy rate of 92\% in detecting LLM-generated fake articles.

However, over the past few years, models' ability to generate text has significantly advanced, raising a bigger challenge.  One recent approach, DetectGPT, introduced by Mitchell et al.~\cite{mitchell2023detectgpt}, aims to address this challenge by detecting whether a given passage is generated by a specific model. The method is based on the hypothesis that AI-generated text exhibits a more negative log probability curvature compared to human-written text. To validate this hypothesis, Mitchell et al. proposed an approximation method for estimating the Hessian trace of the log probability function for both model-generated and human-written text, yielding promising results. However, a limitation of their approach is the requirement of knowing the specific generator model, which may not always be feasible in practice.

In recent research efforts, there has been a specific focus on ChatGPT-generated text due to its global attention. In~\cite{mitrovic2023chatgpt},  Mitrović et al. focused on detecting short texts such as online reviews. They employed a transformer-based model and applied an explanation method (SHAP~\cite{lundberg2017unified}) to gain insights into distinguishing between human-written and machine-generated text. They found that detecting machine-generated text becomes more challenging when it is paraphrased from human text, where a human provides the initial text and asks the model to improve it. Additionally, the authors noted that ChatGPT tends to use uncommon words, exhibits politeness and impersonality, and lacks human-like emotional expressions.
Another related study~\cite{wahle2022large} explored the AI ability in text paraphrasing using GPT-3 and T5 models. They generated machine-paraphrased text and evaluated human performance in detecting these generated texts. The study showed that humans could not accurately detect GPT-3 paraphrased text, with accuracy only slightly above random (53\%).

In response to concerns about AI-generated text, several proprietary tools have emerged to address the detection of AI-authored content. One such tool is GPTZero, which has gained attention in the media as a promising method for identifying AI-generated text~\cite{Svrluga_2023}. However, we were unable to locate the source code or a scientific article detailing their approach. Similarly, numerous applications have been developed claiming to detect AI-generated text, such as GPTkit\footnote{\url{https://gptkit.ai/}}, Illuminarty\footnote{\url{https://illuminarty.ai/en/text/ai-generated-text-detection.html}}, OpenAI's AI Text Classifier\footnote{\url{https://platform.openai.com/ai-text-classifier}}, and AICheatCheck \footnote{\url{https://www.aicheatcheck.com/}}. However, many of these tools lack comprehensive studies on the reliability of their detection methods.

In an effort to facilitate the detection of ChatGPT-generated content, Yu et al.~\cite{yu2023cheat} released a large dataset specifically designed for the identification of ChatGPT-written abstracts. This dataset includes over 35,000 synthetic abstracts generated by ChatGPT, comprising fully generated texts, polished outputs, and mixtures of human-written and machine-generated abstracts. Additionally, the dataset contains more than 15,000 human-written abstracts for comparison. The results of their detection experiments demonstrated the ability to identify content that was entirely generated by ChatGPT. However, the task becomes more challenging when the generated text is mixed with human-written content. This work provides a valuable dataset into the complexities of detecting machine-generated text, particularly in scenarios involving a combination of human and AI-authored content.

A potential solution to address the misuse of LLMs is the use of text watermarks~\cite{grinbaum2022ethical,kirchenbauer2023watermark}. Grinbaum et al.~\cite{grinbaum2022ethical} argue that machine-generated long texts should include a watermark to indicate their source and ensure transparency. In~\cite{kirchenbauer2023watermark}, Kirchenbauer et al. propose embedding watermarks by modifying the sampling rules of next-word prediction. They use a hash function and pseudo-random generator to assign random colors (green and red) to words in the vocabulary. During next-word prediction, words from the red list are prohibited from appearing. However, they acknowledge the difficulty of a watermarking low-entropy text, as substituting a prohibited red word in such cases could result in poor quality output with high perplexity. To address this, they suggest a soft rule that encourages substituting red words in a high-entropy text. 
A third party familiar with the hash function and random generator can easily determine the colors of words by computing them. This detection method does not require knowledge of the specific generation model, making it a cheaper and more straightforward approach.

Although watermarking shows promise as a solution, it is important to consider that it modifies the output text. The method proposed in~\cite{kirchenbauer2023watermark} evaluates quality based on perplexity, but there is a possibility that the meaning and semantics of the output may be altered due to the watermarking process.

\subsection{Challenges and Directions}

Detecting machine-generated text versus human-written text poses increasing challenges as LLMs continue to improve their ability to mimic human language~\cite{baildifficulty}. Several challenges in this regard are highlighted below:

\begin{itemize}
    \item \textbf{Generalization}: Detection methods often lack generalizability, meaning that a method developed to detect text generated by one specific model may not easily transfer to detecting text generated by another model. However, in real-world scenarios, prior knowledge about the specific model generating the text is typically unavailable.
    \item \textbf{Mixed reality}: Existing detection methods struggle when it comes to identifying machine-generated text that is mixed with human-written text. The combination of both types makes it more difficult to differentiate between them.
    \item \textbf{Adaptability}: LLMs demonstrate high adaptability to given prompts, making it challenging for methods that rely on finding patterns in the generated text. Models can exhibit different personalities\footnote{\href{https://learn.microsoft.com/en-us/azure/cognitive-services/openai/how-to/chatgpt?pivots=programming-language-chat-completions\#system-role}{https://learn.microsoft.com/en-us/azure/cognitive-services/openai/how-to/chatgpt}} and respond differently based on prompts, such as ChatGPT's ability to adopt various tones depending on the prompt (e.g., DAN\footnote{\url{https://www.mlyearning.org/dan-chatgpt-prompt/}}), even faking emotions and swear words\footnote{\href{https://metaroids.com/learn/jailbreaking-chatgpt-everything-you-need-to-know/}{https://metaroids.com/learn/jailbreaking-chatgpt-everything-you-need-to-know/}}. This adaptability further complicates detection efforts.
\end{itemize}

Despite these challenges, the remarkable capabilities of LLMs also present great research opportunities for detecting synthetic text. Some potential directions include:

\begin{itemize}
    \item \textbf{Differentiating machine-generated from human-written text}: Achieving this requires collaborative efforts between humans and machines. Humans can contribute their technical knowledge of how models operate and how humans typically express themselves, facilitating the development of detection methods.
    \item \textbf{Attributing the source model of generated text}: Just as humans exhibit distinct writing styles, different language models may possess unique traits when generating text. By differentiating text generated by different models, researchers can gain insights into the behaviors of each model and identify each model's fingerprint.
    \item \textbf{Fact-checking machine-generated text}: LLMs often struggle with generating factual content due to limited training data and the prevalence of fictional stories in their training corpus. In addition, after models are trained, the knowledge stored in these models can quickly become outdated. Conducting fact-checking on machine-generated text is crucial to ensure the reliability of AI-generated information.
\end{itemize}

%% file: sections/metaverse.tex
Neural Radiance Fields (NeRF) have emerged as an effective method for implicit volumetric scene representation, enabling learning from multiple viewing angles~\cite{mildenhall2020nerf}. NeRF has been successfully applied in various domains, including transparent object grasping~\cite{ichnowski2021dex}, scene understanding and reasoning~\cite{zhi2021place,vora2021nesf}, and clear representations in challenging scenarios~\cite{huang2022hdr,guo2022nerfren,ma2022deblur}. Recent NeRF variants, such as NeRF-W~\cite{nerfw} and Ha-NeRF~\cite{hanerf}, have demonstrated their ability to reconstruct scenes from input with various perturbations. We can also see these developments as an example of synthetic realities, especially when we consider the possibility of totally synthesizing new ``worlds''. 

\subsection{NeRF for Metaverse Applications}
In the context of metaverse applications, NeRF has been utilized in virtual concerts~\cite{de2023scannerf} and metaverse platforms for architecture and urban planning~\cite{barron2022mip}. However, challenges remain, such as the need for real-time rendering of complex scenes with multiple dynamic objects and efficient methods to handle large and challenging scenes. Further research is necessary to address these challenges and to explore new use cases for NeRF in metaverse development.

One significant challenge to NeRF's applicability in metaverse development is its reliance on pre-computed camera parameters for scene representation. Several methods, such as NeRF~\cite{nerfmm} and BARF~\cite{lin2021barf}, have been proposed to optimize camera parameters and scene representation. However, avoiding interference from undesired scenes during camera parameter optimization remains an unsolved problem.

A potential research area for the future is building an occlusion-free scene reconstruction based on inaccurate or even unknown camera parameters, enabling greater flexibility in the use of NeRF for scene representation, leading to more effective applications in computer vision and graphics~\cite{mildenhall2020nerf}. In a metaverse, where scenes are typically composed of multiple dynamic objects, avatars, and user interactions, occlusion-free scene reconstruction based on NeRF would enable a more thorough scene representation, resulting in more immersive and realistic virtual environments.

Moreover, an occlusion-free scene reconstruction based on NeRF that does not rely on accurate camera parameters would enable greater flexibility in metaverse development, allowing designers to create and share virtual spaces more efficiently~\cite{barron2022mip}. Finally, optimizing NeRF-based methods for real-time rendering of complex scenes with multiple dynamic objects would enable seamless user interactions in a metaverse, leading to a more responsive and interactive virtual environment~\cite{de2023scannerf}.

In summary, an occlusion-free scene reconstruction based on NeRF has the potential to significantly benefit metaverse applications by enabling more thorough scene representation, flexibility in scene creation and sharing, and real-time performance and scalability. Further research in this area could lead to even more effective applications of NeRF in the context of metaverse development.

\subsection{Challenges and Limitations}
Despite the progress achieved thus far, there are also some challenges and limitations to consider when using NeRF in metaverse development. One significant challenge is the reliance on pre-computed camera parameters for scene representation~\cite{zhu2023occlusion}, making it infeasible when the pre-computation is not possible. This can limit the flexibility of scene creation and sharing. Although some solutions~\cite{nerfmm,lin2021barf} have been proposed to optimize camera parameters along with scene representation, avoiding interference from undesired scenes during camera parameter optimization remains an unsolved problem~\cite{zhu2022neural}.

Another limitation of NeRF is its computational cost. NeRF-based methods require significant computational resources and can suffer from slow rendering times~\cite{muller2022instant}, limiting their real-time performance for complex scenes. This can be a considerable challenge for metaverse applications, where real-time performance is critical for a seamless user experience.

Moreover, NeRF-based methods may not be suitable for all types of scenes. Scenes with complex geometry, occlusions~\cite{zhu2023occlusion}, and dynamic objects~\cite{pumarola2021d} can pose challenges for NeRF-based methods, leading to incomplete scene representation and rendering. Though several image restoration methods~\cite{wang2022low,wan2020reflection,wan2021face,Ma_2019_ICCV,wan2019corrn,wan2022purifying} have been proposed, they are far from being practical solutions for NeRF and its variants. Therefore, it is essential to carefully evaluate the suitability of NeRF-based methods for a given scene and application.

Besides, there are also some potential adverse impacts regarding its broader societal implications.
One potential concern is the potential for NeRF-based metaverse applications to become addictive and negatively impact mental health. The immersive and interactive nature of metaverse environments, combined with the potential for NeRF to create highly realistic and detailed scenes, could create a compelling and addictive experience for users. This could negatively impact mental health~\cite{usmani2022future}, including addiction, social isolation, and other adverse effects associated with prolonged use of virtual environments.

Another potential concern is the impact of NeRF-based metaverse applications on social dynamics and inequality~\cite{lutz2019digital}. NeRF-based metaverse applications could potentially exacerbate existing social inequalities and create new ones. For example, access to high-quality hardware and internet connectivity could become a barrier to participation in these environments, further marginalizing disadvantaged communities.

Last, using NeRF-based metaverse applications has raised significant concerns from both privacy and forensic perspectives~\cite{xing2022ai}. These applications have the potential to collect and store vast amounts of personal data, which could be exploited for targeted advertising, surveillance, and other forms of data mining, leading to further erosion of individual autonomy and privacy. This could result in new forms of digital inequality and harm. Furthermore, the difficulty of collecting and preserving evidence in the NeRF and Metaverse contexts has been discussed in recent literature~\cite{wang2022survey,le2022databases}. Traditional forensic techniques may not be applicable in these virtual environments, where data are decentralized, and ownership is often unclear, making the determination of the chain of custody for digital assets within the Metaverse a complex and challenging task. Additionally, the potential for manipulating digital evidence within these environments raises concerns about the reliability and authenticity of such evidence, particularly with the use of deepfakes and synthetic media~\cite{xing2022ai}. Therefore, there is a pressing need to develop new forensic techniques and tools to address these challenges and ensure the integrity and reliability of digital evidence in the NeRF and Metaverse contexts.

%% file: sections/deepfakes.tex

In the context of synthetic realities, one particular example of utmost attention is deepfakes. Deepfakes are synthetic media that are digitally manipulated to replace one person's identity or personal traits convincingly with that of another. Therefore, when synthetic media comprises the replacement of someone's biometric traits, we are referring to a deepfake. It is typically present in images, audio samples, and videos. 

\subsection{Deepfake Images}
\subsubsection{Deepfake image generation}

The issue of falsified image contents has been a long-standing problem in the image forensics area. With the emergence of deep learning, numerous powerful learning-based models are able to generate the so-called deepfake images with a high level of realism. In recent years, various deepfake techniques have been proposed, including image inpainting/removal, image composition, entire image synthesis, image translation, and text-to-image. Image inpainting/removal is used to fill in image regions with convincing content. Meanwhile, image composition, which encompasses object placement, image blending, image harmonization, and shadow generation, involves cutting out the foreground from one image and pasting it onto another image. Entire image synthesis involves the generation of images entirely by generative models such as GAN~\cite{goodfellow2014generative}, VAE ~\cite{kingma2013auto, kingma2019introduction}, and diffusion models~\cite{sohl2015deep, song2020score}. Image translation, on the other hand, enables the transfer of an image's style, such as converting a sketch image to a colored image. With the rapid development of diffusion models, the images generated based on text prompts are becoming increasingly realistic. Despite their remarkable quality, deepfake images can be misused for malicious purposes, leading to various security issues such as fake news and fraud.

\subsubsection{Deepfake image detection} 
Deepfake image detection methods can be broadly classified into two categories: image-level and pixel-level detection. While image-level methods aim to identify the authenticity of the entire input image, pixel-level methods localize the manipulated regions. Traditional detection methods for detecting image manipulation in image inpainting/removal and image composition, rely on capturing artifacts based on prior knowledge, such as lens distortions~\cite{mayer2018accurate}, CFA artifacts~\cite{ferrara2012image}, noise patterns~\cite{lyu2014exposing}, compression artifacts~\cite{fan2003identification}, etc. Learning-based methods have improved the detection performance by capturing noise prints~\cite{zhou2018learning}, JPEG features~\cite{wang2022jpeg}, High-frequency (HF) artifacts~\cite{zhuo2022self}, and forgery boundary~\cite{dong2022mvss}. Additionally, detecting manipulated images generated through entire image synthesis, image translation, and text-to-image is another challenging problem. Various methods propose to extract visual artifacts~\cite{matern2019exploiting}, color artifacts~\cite{mccloskey2019detecting}, specific GAN fingerprints~\cite{yu2019attributing}, and spectral features~\cite{durall2020watch} for generated image detection. Nevertheless, these methods have limitations in generalizing across different GANs. To address this issue, more general methods such as CNN and generalization methods have been proposed~\cite{wang2020cnn, xuan2019generalization}. As image manipulation technology continues to advance, deepfake image detection is an essential field of research to prevent the spread of misinformation and protect the integrity of visual media.     

\subsubsection{Challenges and future work} 
Despite significant progress in deepfake image detection, there are still several challenges that need to be addressed. One of the main challenges is the generalization of deepfake detection models to unseen datasets and scenarios, which is crucial for practical applications. Another challenge is the robustness of these models against anti-forensics techniques such as recapturing and adversarial attacks. Moreover, the industry is now somewhat ahead of academia in terms of deploying deepfake detection technologies in real-world settings ($e.g.$, ChatGPT). This gap can be narrowed by updating and creating more up-to-date deepfake databases, as most existing ones are somewhat outdated in the research community. Additionally, many deepfake detection models are not explainable, making it difficult to understand how they make decisions. Future work should aim to develop explainable models that can provide clear and interpretable justifications for their decisions. Overall, addressing these challenges can lead to more reliable and effective deepfake image detection systems in the future.



\subsection{Deepfake Video}
\subsubsection{Deepfake video generation}
Recently, deepfake videos typically refer to manipulated face videos. Face information plays a vital role in human communication~\cite{frith2009role}. However, the spread of deepfake videos on social media platforms can result in significant security concerns due to the potential dissemination of disinformation and misinformation, posing tangible and pressing security concerns.  
Generally speaking, there are four primary categories of deepfake videos, which are identity swap, face reenactment, attribute manipulation, and entire synthesis~\cite{kong2022digital}. These videos are generated using powerful generative models such as GAN~\cite{goodfellow2014generative}, VAE~\cite{kingma2013auto, kingma2019introduction}, and diffusion models~\cite{sohl2015deep, song2020score}, which are capable of producing highly sophisticated videos. Identity swap replaces the original face regions with target faces, while face reenactment transfers the source facial expression to the target one. Attribute manipulation can alter specific facial features like hair, eyeglasses, nose, etc. With the advent of foundation models, entire synthesized videos can be generated. Powerful deep learning tools have been used to create sophisticated deepfake video datasets like UADFV \cite{yang2019exposing}, DF-TIMIT~\cite{korshunov2018deepfakes}, FaceForensics++~\cite{rossler2019faceforensics++}, DFD~\cite{dufour2019contributing}, DFDC~\cite{dolhansky2019deepfake}, Celeb-DF~\cite{li2020celeb}, DF-Forensics-1.0~\cite{jiang2020deeperforensics}, ForgeryNet~\cite{he2021forgerynet}, FFIW~\cite{zhou2021face}, KoDF~\cite{kwon2021kodf}, and FakeAVCeleb~\cite{khalid2021fakeavceleb}. As deepfake techniques continue to evolve, it is crucial to develop effective methods for detecting deepfake videos and preventing their malicious use.

\subsubsection{Deepfake video detection} 
To counteract malicious deepfake attacks, many detection methods have been proposed. Traditional methods mainly focus on hand-crafted features, such as lack of eye-blinking~\cite{li2018ictu} and warping artifacts~\cite{li2018exposing}. However, these methods are not accurate enough. Learning-based methods such as 
convolutional neural networks (CNN)~\cite{afchar2018mesonet, nguyen2019capsule, chollet2017xception, tan2019efficientnet}, recurrent neural networks (RNN)~\cite{sabir2019recurrent}, and vision transformer (ViT)~\cite{heo2021deepfake}, have been proposed to achieve more promising detection performance. Afchar $et~al.$ \cite{afchar2018mesonet} designed MesoNet and MesoInception4 to detect Deepfake and Face2Face videos automatically. Besides, some generic networks such as Xception Net~\cite{chollet2017xception}, Efficient Net~\cite{tan2019efficientnet}, and Capsule Net~\cite{nguyen2019capsule} have been demonstrated effective on deepfake detection tasks. Subsequent works have employed RNN \cite{sabir2019recurrent} and ViT \cite{heo2021deepfake} to further improve forgery detection accuracy. Other methods capture spatial artifacts~\cite{chen2021local, shang2021prrnet,nirkin2021deepfake,li2020sharp, zhao2021multi, kumar2020detecting}, frequency artifacts~\cite{qian2020thinking,miao2022hierarchical, kong2022detect}, and biological signals~\cite{qi2020deeprhythm, ciftci2020fakecatcher} to perform deepfake detection. Follow-up works~\cite{li2020face, luo2021generalizing, liu2021spatial, shiohara2022detecting, zhao2021learning, cao2022end, zhu2021face} focus on improving the generalization capability and robustness of the model. Temporal information has also been exploited in many deepfake video detection methods based on typical generic networks such as 3DCNN~\cite{zhang2021detecting}, LSTM~\cite{amerini2020exploiting, hochreiter1997long}, RNN~\cite{chintha2020recurrent, sabir2019recurrent}, and ViT~\cite{khan2021video}. Combining spatial and temporal information can achieve more reliable detection and improve the model's generalization capability.
 
\subsubsection{Challenges and future work} 
Deepfake video creation and detection have seen significant success in recent years, but many issues remain unresolved. While accurate and secure, deepfake detectors lack interpretability, limiting their applications in practical scenarios. Localizing forgery regions and forgery frames is also a crucial yet understudied task. Additionally, the two-player nature of face forgery and forgery detection means that attack techniques will continue to become more powerful, thereby calling for more general detection methods. Furthermore, deepfake videos often involve audio manipulation, which is largely overlooked in existing methods. Therefore, more visual-audio joint datasets and multi-modal detectors are expected in future works.

\subsection{Deepfake Audio}
\subsubsection{Deepfake audio generation}

Deepfake audio refers to manipulated or synthetic audio created using deep learning techniques. The aim of deepfake audio is to impersonate the speaker's speech characteristics, such as accent, timbre, and intonation, by learning from target voice resources. Traditional methods for audio manipulation involve removing, duplicating, copying within an audio sample, or pasting and inserting fragments into other audios. Deep learning-based speech synthesis makes the generated audio more realistic and difficult to distinguish from real ones. Subsequent models based on likelihood algorithms, such as WaveNet~\cite{oord2016wavenet} and WaveGlow~\cite{prenger2019waveglow}, have been developed to perform audio generation. However, these methods often require conditional information and may fail to generate long signal sequences. Recent waveform generative models, such as GAN~\cite{kumar2019melgan, donahue2020end, yamamoto2020parallel} and VAE~\cite{peng2020non}, take advantage of various auxiliary losses, thereby achieving superior generation performance. On the other hand, recent diffusion models ($e.g.$, DiffWave~\cite{kong2020diffwave}), have exhibited remarkable generation performance even in challenging unconditional and class-conditional waveform generation scenarios. In the context of text-to-speech tasks, diffusion models can be classified into: acoustic model ($e.g.$, Diff-TTS~\cite{jeong2021diff}), vocoder ($e.g.$, DiffWave~\cite{kong2020diffwave}), and end-to-end framework ($e.g.$, FastDiff~\cite{huang2022fastdiff}). The promising results of diffusion models indicate their potential to revolutionize the field of audio generation and synthesis.

\subsubsection{Deepfake audio detection}

Automatic Speaker Verification (ASV) systems~\cite{naika2018overview} currently detect manipulated audio through three tasks: logical access (LA), physical access (PA), and speech deepfake (DF)~\cite{yamagishi2021asvspoof}. The LA task focuses on detecting synthetic speech injected into a communication system, while the PA task includes acoustic propagation and real physical factors. The DF task aims to detect deepfake speech circulating on social media platforms. Traditional methods for detecting deepfake audio involve analyzing the spectrogram of the audio and exposing audio inconsistencies, such as abrupt changes in frequency or amplitude. These methods are based upon the assumption that synthetic audios have unique frequency and amplitude patterns. However, recent deep learning deepfake techniques raise the difficulty in identifying authenticity and call for the design of deep learning countermeasures. Generally speaking, deep learning methods can be categorized into feature-based, image-based, and waveform-based~\cite{cai2023waveform,sun2023ai}. Feature-based methods utilize critical digital signal features, such as Mel-frequency cepstral coefficients (MFCCs)~\cite{logan2000mel}, constant Q cepstral coefficient (CQCC)~\cite{yang2018extended}, and energy, to detect deepfake audio. Image-based methods apply the spectrogram image of the signal to conduct inconsistency detection. Waveform-based methods aim to analyze the raw waveform of the audio signal instead.
To facilitate the development of deepfake audio detection models, numerous databases, such as M-AILABS Speech~\cite{M-AILabs}, GAN based synthesized audio dataset~\cite{dat9-0j82-19}, Half-Truth~\cite{yi2021half}, and H-Voice~\cite{ballesteros2020dataset} have been created. Overall, deepfake audio detection is a challenging but important task that requires the constant development and refinement of detection methods. 

\subsubsection{Challenges and future work} 

Despite recent advancements in synthetic audio generation, there are still several challenges that need to be addressed. As generation techniques continue to evolve, deepfake audio will become increasingly difficult to distinguish by both human and AI-based detectors. Even worse, existing detection methods have shown poor robustness to compression, encoding, and noise. Additionally, current deepfake audio detection methods suffer from inefficient training datasets and overfitting issues~\cite{yamagishi2021asvspoof}, resulting in limited generalization capability. Moreover, most detection methods extract specific features (such as MFCC, CQCC, and, energy) to conduct deepfake detection. However, it is challenging to extract appropriate features for specific detection tasks. How to effectively combine various features for more robust detection opens an important research path forward. Last but not least, it is crucial to conduct further research on the ethical implications of deepfake audio, such as its potential for misuse and its impacts on audio professionals.

%% file: sections/challenges.tex


Trust plays a fundamental role in our society. 
Citizens entrust infrastructures, services (including education and health), media (including social networks nowadays), the judiciary system (including law enforcement), and political decision-making in general. 
Democracies are endangered when citizens no longer trust the system and their elected representatives.

Unfortunately, trust can be tampered with through influence or disinformation. 
Although disinformation has probably always existed in human history (e.g., spreading rumors to influence elections), the message's quality and scale were low, restricting its impact. However, in our contemporary digital world, disinformation (also coined fake news) with greater realistic content spreads at an unprecedented scale on the Internet through alternative media without any filtering by the traditional mainstream channels. With the advance of Artificial Intelligence (AI) technologies, all sorts of media (text, images, and audio) can be synthetically generated. More particularly, recent generative AI models trained on very large datasets can produce more plausible and realistic content.

Distinguishing truth from falsity is becoming even more difficult, and the difference between reality and fiction is getting thinner daily.
We are now facing the \textbf{Era of Synthetic Realities}.
Disinformation exploits cognitive biases (e.g., anchoring bias, third-person effect, authority bias, bandwagon effect, to mention a few)~\cite{kahneman_2017}, which are systematic errors in judgment that humans can make, and because of this, synthetic realities represent a threat to our society. 

Synthetic realities are now generated by criminals and hostile agents for various malicious operations, including:
political disinformation and state espionage (e.g., fake social network profiles),
national security (e.g., facilitating a military coup),
financial fraud (e.g., CEO scam impersonation),
blackmail (e.g., ransomfake),
defamation (e.g., revengeporn),
plausible deniability of Forensic evidence.


%
As human beings, because of cognitive biases, citizens will never stop falling for disinformation. A way to fight head on is to create tools to analyze digital content prior authentication by human experts. We anticipate some factors of utmost importance when developing new solutions. 

The first one involves exploring the context of a digital asset as much as possible, even with limited training data. The second one involves efforts on robustness and interpretability, as decisions must be intelligible to human beings. The final one consists in being conscious of the incompleteness of individual methods and orchestrating decision-making fusion methods to combine different telltales for final detection. 
\begin{enumerate}

\item \textbf{Limited training data}. Data-driven approaches often rely upon large amounts of training data. This problem may become critical in the rapidly evolving scenario of fake information. New forms of falsification, unknown to the forensic analyst, are proposed daily, preventing the timely collection of all relevant training data. We discuss this issue by posing the problem as an open-set recognition problem~\cite{scheirer2012toward}; that is, we need to define a suitable model for pristine data and analyze false information by looking for inconsistencies concerning this model. If possible, researchers also need to consider few-shot learning approaches that only require a tiny amount of labeled data to update to new threats.
\item \textbf{Robustness and Interpretability}. Being robust to wide-spectrum and unforeseen conditions is a basic system requirement, but it becomes central in a forensic environment featuring two active players. Robustness to adversarial attacks is paramount nowadays, especially when data-driven methods are applied, in light of the many literature findings that emphasize their vulnerability. Besides using basic solutions, we pose that reliability by relying on interpretable machine learning is necessary. We advocate for strategies that help us understand why a learning-based system behaves a certain way and provides the observed answers. Methods should also include semantics and context to support the entire decision-making process.
\item \textbf{Fusion}. Combining different methods toward a unified detection framework is very promising. As discussed in prior art for image forgery detection~\cite{ferreira2016behavior}, fusion in different learning stages (early, middle, or even late-stage) plays a fundamental role within dynamic and adversarial setups. We envision learning strategies combining different telltales as a promising way forward. 
\end{enumerate}

Therefore, some driving research questions involve challenges in:
\begin{itemize}
    \item \textbf{Detection}: is it possible to detect plausible and realistic digital content (e.g., text generated by large-scale language models, synthetic images, and voices generated by generative models)? 
    \item \textbf{Attribution}: is it possible to accomplish source attribution by assigning manipulated digital content to a known type of attack vector? 
    \item \textbf{Explainability}: is it possible to automatically uncover cues or inconsistencies in digital content to corroborate falsity, as discussed above? 
    \item \textbf{Context and fusion}: how to incorporate context? How to combine different telltales?
\end{itemize}

Many other challenges will play out in the coming years as synthetic realities become ever more realistic, directly affecting fundamental pillars of our society, such as democratic values, individual freedom, and social tolerance. In this paper, we strived to discuss some of these challenges and what lies ahead, but it was just the tip of the iceberg. 

Only an orchestrated effort of government representatives, society at large, and researchers will be able to curb such threats. We believe possible explorations might lie in regulatory acts, education investments, and scientific research for more powerful detection methods.